\documentclass[12pt,a4paper]{article}
\begin{document}
\thispagestyle{empty}

\begin{flushright}
UAB--FT/97--412\\
hep-ph/yymmxxx\\
March 1997
\end{flushright}
\vspace{1.5cm}

\begin{center}
\noindent {\bf MIXING OF $\eta$-$\eta\prime$ MESONS IN $J/\psi$ DECAYS INTO \\
     A VECTOR AND A PSEUDOSCALAR MESON} \\
\vspace{2.5cm}

A. BRAMON$^a$, R. ESCRIBANO$^{a,b}$ and M. D. SCADRON$^c$
\vspace{0.8cm}\\

$^a$Departament de F{\'\i}sica and $^b$Institut de F{\'\i}sica d'Altes Energies, \\
Universitat Aut\`onoma de Barcelona, E-08193 Bellaterra (Barcelona), Spain \\
\vspace{0.2cm}

$^c$Physics Department, University of Arizona, \\ Tucson, AZ 85721, USA \\ 
\vspace{2.5cm}

{\bf ABSTRACT}
\end{center}

The octet-singlet mixing angle $\theta_P$ in the pseudoscalar meson nonet is deduced
from the rich set of accurate data on $J/\psi$ decays into a vector and a
pseudoscalar meson. Corrections due to non-ideal $\omega$-$\phi$ mixing have been
included for the first time  and turn out to be crucial to find 
$\theta_P = -16.9^\circ \pm 1.7^\circ$, 
which is appreciably less negative than previous results coming from similar analyses.

\def\tableone{
\begin{table}
\centering
\caption{Experimental $J/\psi \rightarrow VP$ branching ratios from PDG \cite{PDG} and
results of our fits. $BR$'s for all $VP$ channels are in $10^{-3}$.}
\label{table1}
\vspace*{0.5cm}
\begin{tabular}{|c|c|c|c|}
\hline
                       &                    &                      &
\\[-1.0ex]  
                       & Exp.~Ref.~\cite{PDG}              
& Fit (Ref.~\cite{COF})        & Fit (Ref.~\cite{JOU})
\\[1.0ex]
\hline
                       &                    &                      &
\\[-1.0ex]
$\rho \pi$             & 12.8   $\pm$ 1.0   & 11.1    $\pm$ 0.8    & 11.2    $\pm$
0.8    \\[0.5ex]
$K^{\ast +} K^-  
+ c.c.$                &  5.0   $\pm$ 0.4   &  5.17   $\pm$ 0.51   &  5.05   $\pm$
0.48   \\[0.5ex]
$K^{\ast 0} {\bar K}^0 
+ c.c.$                &  4.2   $\pm$ 0.4   &  4.41   $\pm$ 0.56   &  4.48   $\pm$
0.55   \\[0.5ex]
$\omega \eta$          &  1.58  $\pm$ 0.16  &  1.71   $\pm$ 0.14   &  1.69   $\pm$
0.14   \\[0.5ex]
$\omega \eta'$         &  0.167 $\pm$ 0.025 &  0.173  $\pm$ 0.030  &  0.177  $\pm$
0.031  \\[0.5ex]
$\phi \eta$            &  0.65  $\pm$ 0.07  &  0.665  $\pm$ 0.122  &  0.671  $\pm$
0.115  \\[0.5ex]
$\phi \eta'$           &  0.33  $\pm$ 0.04  &  0.305  $\pm$ 0.086  &  0.278  $\pm$
0.074  \\[0.5ex]
$\rho \eta$            &  0.193 $\pm$ 0.023 &  0.212  $\pm$ 0.019  &  0.209  $\pm$
0.019  \\[0.5ex]
$\rho \eta'$           &  0.105 $\pm$ 0.018 &  0.0963 $\pm$ 0.0079 &  0.0996 $\pm$
0.0081 \\[0.5ex]
$\omega \pi^0$         &  0.42  $\pm$ 0.06  &  0.373  $\pm$ 0.031  &  0.374  $\pm$
0.031  \\[0.5ex]
$\phi \pi^0$           &$<0.0068$           &  0.0011 $\pm$ 0.0001 &  0.0011 $\pm$
0.0001
\\[1.0ex]
\hline
                       &                    &                      &
\\[-1.0ex]
$g$                  & & 1.065  $\pm$ 0.036 & 1.075  $\pm$ 0.038   \\[0.5ex]
$s$                  & & 0.097  $\pm$ 0.031 & 0.112  $\pm$ 0.027   \\[0.5ex]
$e$                  & & 0.117  $\pm$ 0.005 & 0.117  $\pm$ 0.005   \\[0.5ex]
$\theta_e$           & & 1.29   $\pm$ 0.16  & 1.35   $\pm$ 0.16    \\[0.5ex]
$r$                  & & -0.148 $\pm$ 0.009 & -0.151 $\pm$ 0.009   \\[0.5ex]
$X_\eta$             & & 0.794  $\pm$ 0.014 &  0.786 $\pm$ 0.014   \\[1.0ex]
\hline
\end{tabular}
\end{table}
}

\newpage

The value of the $\eta$-$\eta\prime$ mixing angle $\theta_P$ in the
pseudoscalar-meson nonet has been discussed many times in the last thirty years. A
well-known contribution to this discussion is the  phenomenological analysis
performed by Gilman and Kauffman \cite{GK} almost a decade ago.  The approximate
value $\theta_P \simeq -20^\circ$ was proposed by these authors through  a rather
exhaustive discussion of the experimental evidence available at that time. Another,
more recent analysis by two of the present authors \cite{BS} concluded that a
somewhat less negative value, $\theta_P = -14^\circ \pm 2 ^\circ$, seems to be
favoured. A significant difference between these two independent analyses concerns
the set of rich data on $J/\psi$ decays into a vector and a pseudoscalar meson,
$J/\psi \to V P$, which were included in the first analysis
\cite{GK} but not in the second one \cite{BS}. The purpose of the present note is to
extract a value for $\theta_P$ from this relevant set of $J/\psi \to V P$ decay data.
In this sense, we will follow quite closely the similar analyses in 
Refs.~\cite{COF,JOU,MOR} except that the apparently negligible effects of non-ideal
mixing in the vector-meson nonet will be now taken into account. Ignoring these
effects, {\it i.e.}, assuming that the physical $\omega$ and $\phi$ coincide precisely
with the ideally mixed states $\omega_0 \equiv (u {\bar u} + d {\bar d})/ \sqrt2$ and
$\phi_0 \equiv  s {\bar s}$, lead to $\theta_P = -19.2^\circ \pm 1.4^\circ$
(Ref.~\cite{COF}), 
$\theta_P = -19.1^\circ \pm 1.4^\circ$ (Ref.~\cite{JOU}) and $\theta_P \simeq
-20^\circ$ (Ref.~\cite{MOR}). Introducing the small, but certainly non-vanishing,
departure of $\omega$ and $\phi$ from the above ideally mixed states $\omega_0$ and 
$\phi_0$, and using essentially the same set of $J/\psi \to V P$ data as in 
Refs.~\cite{COF,JOU,MOR}, we will obtain $\theta_P = - 16.9^\circ \pm 1.7^\circ$.

Experimental information concerning strong and electromagnetic 
$J/\psi$ decays into a vector and a pseudoscalar meson, 
$J/\psi \to V P$ with $V = \rho, \omega, \phi$ or $K^*$ and $P = \pi, \eta,
\eta\prime$ or $K$,  comes mainly from the MARK III \cite{COF} and DM2 \cite{JOU}
detectors.  We have collected all this information on the $J/\psi \to V P$ branching
ratios ($BR$), as averaged by the PDG compilation \cite{PDG}, in the first column of
Table \ref{table1}. The highly accurate value for $BR(\rho
\pi)$ comes from  Ref.~\cite{COF} and other papers listed in \cite{PDG}; the upper
limit for $BR(\phi \pi)$ has been established by \cite{COF}; and the nine remaining
$BR$'s, with relative experimental errors ranging from about 8 to 17 \%, come from
Refs.~\cite{COF} and \cite{JOU}. Altogether they constitute an excellent and
exhaustive set of data which remains unchanged in the recent editions of the PDG
compilations. Part of these data were already used in the analyses of 
Refs.~\cite{COF,JOU,MOR}; our purpose here consists also in improving these analyses
by using the complete set.

Attempts to understand these decays in a phenomenological context star\-ted
immediately after the appearence of the data and all these attempts were based on
the same essential model with slight variations \cite{COF,JOU,MOR,BC}. The dominant
piece of the amplitude is unanimously assumed to proceed through the annihilation of
the initial
$c \bar c$ pair into the $SU(3)$-flavorless part of the final $VP$ system via three
(or more) gluons; we will denote this strong interaction piece of the amplitude by
$g$. The non-vanishing of the 
$BR(\omega \pi)$ and the differences between the $BR$'s into charged or neutral $K^*
{\bar K} +  K {\bar K^*}$  systems, clearly requires an electromagnetic piece in the
amplitude coupling to both the isoscalar and isovector parts of the final $VP$; this
correction to the dominant part of the amplitude will be denoted by $e$ (the phase
of $e$ relative to $g$ is defined to be $\theta_e$).  Apart  from these two
contributions, associated to ``connected'' diagrams, a good fit is achieved only if
``disconnected'' (Ref.~\cite{COF}) or, equivalently, ``doubly-OZI-violating''
(Ref.~\cite{JOU}) diagrams are introduced too; their contribution to the amplitude
will be denoted by
$r g$, with $r<1$ being the ratio between this latter correction and the dominant
piece $g$. The explicit amplitudes of Ref.~\cite{JOU} are then easily obtained (with
our $g s$ terms called
$h$ in Ref.~\cite{JOU}),
\begin{equation}
\label{amp}
\begin{array}{lcl}
A(\rho\pi) &=& g + e\ , \\[0.5ex] 
A(K^{*\pm} K^{\mp}) &=& g(1-s) + e(2-x)\ , \\[0.5ex] 
A(K^{*0} \bar K^0) &=& g(1-s) - 2e(1+x)/2\ , \\[0.5ex] 
A(\omega_0 \eta) &=&  (g+e) X_{\eta} + 
                 \sqrt2 r g (\sqrt2 X_{\eta}+Y_{\eta})\ , \\[0.5ex]
A(\omega_0 \eta\prime) &=& (g+e) X_{\eta\prime} + 
                       \sqrt2 rg (\sqrt2 X_{\eta\prime}+Y_{\eta\prime})\ , \\[0.5ex] 
A(\phi_0 \eta) &=& [g(1-2s)-2ex] Y_{\eta} + 
               r g(1-s)(\sqrt2X_{\eta}+Y_{\eta})\ , \\[0.5ex] 
A(\phi_0 \eta\prime) &=& [g(1-2s)-2ex] Y_{\eta\prime} + 
                     r g(1-s)(\sqrt2X_{\eta\prime}+Y_{\eta\prime})\ , \\[0.5ex]   
A(\rho\eta) &=& 3 e X_{\eta}\ , \\[0.5ex] 
A(\rho \eta\prime) &=& 3 e X_{\eta\prime}\ , \\[0.5ex] 
A(\omega_0 \pi^0) &=&  3 e\ , \\[0.5ex]  
A(\phi_0 \pi^0) &=&  0\ ,
\end{array}
\end{equation}
where $X_{\eta} = Y_{\eta\prime} = \cos \phi_P$ and $X_{\eta\prime} =
- Y_{\eta} = \sin \phi_P$, and the first three amplitudes include all final particle
charge states. 
The two $\eta$-$\eta\prime$ mixing angles, $\theta_P$ and
$\phi_P \equiv \theta_P + 54.7^\circ$, refer to the octet-singlet and
non-strange--strange basis, respectively,
\begin{equation}
\label{mix}
\begin{array}{rcl}
\eta &\equiv& \cos \theta_P \eta_8 - \sin \theta_P \eta_1 
\equiv \cos \phi_P (u \bar u + d \bar d)/ \sqrt 2 - \sin \phi_P s \bar s\ , \\[0.5ex]
\eta\prime &\equiv& \sin \theta_P \eta_8 + \cos \theta_P \eta_1 
\equiv \sin \phi_P (u \bar u + d \bar d)/ \sqrt 2 + \cos \phi_P s \bar s\ . 
\end{array}
\end{equation} 
The parameters $s$ and $x$ in Eq.~(\ref{amp}) account for
$SU(3)$-breaking corrections penalizing the creation of strange quarks over
non-strange ones; $s$ is left as a free parameter, but $x$ is fixed to $x=0.64$ in
the DM2 analysis \cite{JOU}.
This latter value can be qualitatively understood as the ratio of non-strange to
strange constituent quark masses arising from the quark 
propagators \cite{JS,S}. 
Alternatively, one can fix $x$ to its $SU(3)$-symmetric value,
$x=1$, as in the MARK III analysis \cite{COF}. 
The amplitudes used in Ref.~\cite{COF} follow from Eq.~(\ref{amp}) by simply fixing
$x=1$ and ignoring (second order) correction terms proportional to
$s r$. The differences in our results induced by this different treatment of
non-leading corrections will be later used to estimate the theoretical uncertainties
associated to the model.

We have now the possibility of performing the fits corresponding to the slightly
different  models in Refs.~\cite{COF} and \cite{JOU} using now the whole set of
data.  The various branching ratios,
$BR(VP)$, are simply related to the corresponding amplitude in Eq.~(\ref{amp}) by 
$BR(VP) = |A(VP)|^2 |\vec {p_V}|^3$, where $BR(VP)$ is given in $10^{-3}$ if the
$V$-meson momentum in the phase-space factor is given in $GeV$. Proceeding precisely
either as in Ref.~\cite{COF} or as in Ref.~\cite{JOU}, our fit to the whole set of
data leads to
\begin{equation}
\label{old}
\begin{array}{rcl}
\theta_P &=& -19.6^\circ \pm 1.4^\circ \ (x=1)\ , \\[0.5ex] 
\theta_P &=& -18.9^\circ \pm 1.4^\circ \ (x=0.64)\ ,
\end{array}
\end{equation}
with $\chi^2$-values for 4 degrees of freedom (d.o.f.) of $7.3$ or $6.2$,
respectively. Our results for the mixing angle are therefore quite consistent with
those coming from the two previous analyses leading to $\theta_P = -19.2^\circ \pm
1.4^\circ$ \cite{COF} and 
$\theta_P = -19.1^\circ \pm 1.4^\circ$ \cite{JOU}, where different subsets of the
presently available (and consistent) data had been used.  We can also attempt a
global fit with the amplitudes proposed in Ref.~\cite{MOR} which differ from those in
Eq.~(\ref{amp}) only in non-leading $s r$ terms and the value of $x$ ($x=0.70$).  We
obtain $\theta_P = -19.4^\circ \pm 1.4^\circ$ in agreement with 
$\theta_P = -20.2^\circ$ from Ref.~\cite{MOR} (Solution II) but with a poor $\chi^2 =
12.3$ for 4 d.o.f. A similar value, $\theta_P \simeq -19^\circ $, was obtained in 
Ref.~\cite{BC}. Averaging the values in Eq.~(\ref{old}) and estimating from their
difference (or from this whole discussion) the uncertainties in the model, one can
safely conclude that
$J/\psi \to V P$ data favour the value 
$\theta_P = -19.3^\circ \pm 1.7^\circ$ for the $\eta$-$\eta\prime$ mixing angle {\it
if} $\omega$ and $\phi$ are assumed to coincide with the ideally mixed states
$\omega_0$ and $\phi_0$.

The possibility of improving our previous fits and modifying our latter value for the 
$\eta$-$\eta\prime$ mixing angle is apparent if one simply restricts the analysis to
the subset of data in Table \ref{table1} which do not involve $\omega$ and $\phi$.
For instance, from the ratio between 
$BR(\rho \eta\prime)$ and $BR(\rho \eta)$, which is proportional to $\tan^2 \phi_P$,
one immediately obtains $\phi_P = 40.0^\circ \pm 3.3^\circ$ or, equivalently,
$\theta_P = - 14.7^\circ \pm 3.3^\circ$. 
Similarly, performing a partial fit to the six $BR$'s with $V \neq \omega, \phi$
leads to $\theta_P = -14.5^\circ \pm 2.8^\circ$ (with $x=1$, as in
\cite{COF}, and $\chi^2 = 0.62$ for one d.o.f.) and to the same value $\theta_P =
-14.5^\circ \pm 2.8^\circ $ (with $x=0.64$, as in \cite{JOU}, and
$\chi^2 = 0.62$ for one d.o.f.). In all these cases, the value for $\theta_P$ turns
out to be considerably less negative than the values in Eq.~(\ref{old}). This strongly
suggests that a more detailed analysis incorporating the small, but unambiguously
established, $\omega$-$\phi$ mixing effects can be particularly relevant. The rest of
the present note is devoted to this kind of analysis.

The departure from $\omega$-$\phi$ ``ideal'' mixing is usually parametrized by the
small angle $\phi_V$ defined through
\begin{equation}
\label{phi}
\begin{array}{rcl}
\omega \equiv \cos \phi_V \omega_0 - \sin \phi_V \phi_0\ ,\\[0.5ex] 
\phi   \equiv \sin \phi_V \omega_0 + \cos \phi_V \phi_0\ ,
\end{array}
\end{equation} 
where from now on $\omega$ and $\phi$ refer to the physical states and
$\omega_0$ and $\phi_0$ refer to the ideally mixed states $(u\bar u +d\bar d)/\sqrt2$
and $s \bar s$ used so far. The value of the $\omega$-$\phi$ mixing angle is well
known and will be fixed to 
\begin{equation}
\label{sinphiV}
\sin \phi_V \simeq \tan \phi_V = +0.059 \pm 0.004\ ,
\end{equation}
or $\phi_V \simeq + 3.4^\circ$. The modulus of $\tan \phi_V$ can be obtained via the
well-known and clearly understood ratio \cite{PDG,JS}
\begin{equation}
\Gamma (\phi\to \pi^0\gamma)/\Gamma(\omega\to\pi^0\gamma) = 
\tan^2 \phi_V |p_\phi / p_\omega|^3 = (8.10\pm 0.94)\times 10^{-3}\ ,
\end{equation}
and is fully compatible with the values coming from the squared Gell-Mann--Okubo mass
formula  ($\phi_V\sim 39^\circ - 35.3^\circ \sim +3.7^\circ$, see Ref.~\cite{PDG})
and from $\omega$-$\phi$ interference in $e^+e^- \to \pi^+\pi^-\pi^0$ data (see
Ref.~\cite{BGP}). Apart from confirming the modulus of $\tan \phi_V$, these latter
data fix unambiguously its phase too. 
 
An improved description of the $J/\psi$ decay amplitudes into $P=\eta,\eta\prime$ and 
$V=\omega,\phi$ can now be immediately obtained from Eqs.~(\ref{amp}), (\ref{phi}) and
(\ref{sinphiV}), and the corresponding fits to the same global set of data can be
performed as before. The results of these fits are shown in the second and third
columns of Table \ref{table1}. For the $\eta$-$\eta\prime$ mixing angle one now
obtains
\begin{equation}
\begin{array}{rcl}
\label{new}
\theta_P &=& - 17.3^\circ \pm 1.3^\circ \ (x=1)\ ,\\[0.5ex] 
\theta_P &=& - 16.6^\circ \pm 1.3^\circ \ (x=0.64)\ ,
\end{array}
\end{equation}
which are values significantly less negative (by almost 2 standard
deviations) than the corresponding ones, Eqs.~(\ref{old}), obtained neglecting
$\omega$-$\phi$ mixing effects. The quality of the fits has also been slightly
improved to $\chi^2 =6.0$ and $5.4$ for the same four d.o.f. as before; indeed, the
only difference between the fits leading to (\ref{old}) or to (\ref{new}) is having
fixed the value of $\phi_V$ either to $0$ or to $3.4$ degrees. If the value of
$\phi_V$ is left as a free parameter to be fixed through $J/\psi \to V P$ decay data,
one then obtains 
$\theta_P=-16.9^\circ\pm 2.6^\circ,\; \phi_V=4.0^\circ\pm 3.2^\circ$ and
$\chi^2=6.0/3$ d.o.f. (for $x=1$, Ref.~\cite{COF}) or 
$\theta_P=-16.8^\circ\pm 2.7^\circ,\; \phi_V=3.1^\circ\pm 3.2^\circ$ and
$\chi^2=5.4/3$ d.o.f. ($x=0.64$, Ref.~\cite{JOU}). These latter results suggest that
in the set of $J/\psi \to V P$ decay data {\it alone} there are indications of
slightly non-ideal $\omega$-$\phi$ mixing compatible with Eq.~(\ref{sinphiV}) thus
confirming our whole approach. Obviously, the previously discussed information on
$\phi_V$, Eq.~(\ref{sinphiV}), coming from non-$J/\psi$ physics has to be taken into
account too, as we did when obtaining our values for
$\theta_P$ in Eqs.~(\ref{new}). Taking their average and estimating the theoretical
error as before leads to
\begin{equation} 
\label{fin} 
\theta_P = - 16.9^\circ \pm 1.7^\circ\ ,
\end{equation}
which seems to us the most reasonable value for the $\eta$-$\eta\prime$
mixing angle that one can extract from the available $J/\psi \to V P$ decay data.
 
Our final 
value $\theta_P = -16.9^\circ \pm 1.7^\circ$ is somehow in between the values 
$\theta_P \simeq -20^\circ$ and $\theta_P = -14^\circ \pm 2^\circ$ obtained in the
old but more general analyses in Refs.~\cite{GK,BS}. A similarly exhaustive analysis,
including all the relevant experimental information, is presently in progress. A
recent discussion involving  several channels concludes independently that
$\theta_P$ can range between $-20$ and $-17$ degrees \cite{BALL}. 
Also, the dedicated analysis by the Crystal Barrel Collaboration
\cite{AMS} (not discussed by the previous authors) favours $\theta_P = -17.3^\circ\pm
1.8^\circ$, quite in line with our final result (\ref{fin}). A further confirmation
comes from the recent analysis of semileptonic $D_s$ decays \cite{MEL} favouring a
mixing angle in the range $-18^\circ \leq \theta_P \leq -10^\circ$ with the best
agreement observed for $\theta_P = -14^\circ$. However, a crucial test for the value
of $\theta_P$ is expected to come in a near future from DA$\Phi$NE, where the ratio
$R_\phi \equiv
\Gamma(\phi\to\eta\prime\gamma)/
\Gamma(\phi\to\eta\gamma)$ will be accurately measured; an analysis following 
Ref.~\cite{BS} predicts $R_\phi = 7.6 \times 10^{-3}$ for $\theta_P = -20^\circ$ and 
$R_\phi = 6.2 \times 10^{-3}$ for our slightly less negative value 
$\theta_P = -16.9^\circ$. Our prediction for the branching ratio $J/\psi \rightarrow
\phi\pi^0$ are not much below the available experimental upper limit (see Table
\ref{table1}). A measurement of this $BR(\phi\pi^0)$ would also be crucial to confirm
or falsify our approach.

One can use our value for $\theta_P$ to estimate the parameter $\Delta$ (see
Ref.~\cite{PDG}) or $\Delta_M$ (see Ref.~\cite{LEU}) accounting for the violation of
the Gell-Mann--Okubo mass formula,
\begin{equation}
m_{88}^2 = \frac{1}{3} (4 m_K^2 - m_{\pi}^2) \{1+\Delta\} \equiv
           \frac{1}{3} (4 m_K^2 - m_{\pi}^2) + 
           \frac{4}{3} (  m_K^2 - m_{\pi}^2) \Delta_M\ .
\end{equation}
From these Eqs., our result (\ref{fin}) and the well-known relation 
$\theta_P = -10.1^\circ (1+8.5 \Delta)$ \cite{PDG}, one obtains 
$\Delta_M = 0.074 \pm 0.019$ quite in line with $\Delta_M = 0.065 \pm 0.065$ coming
from the recent analysis in Ref.~\cite{LEU}. Moreover, from the
discussion in Ref.~\cite{LEU2} one has
\begin{equation}
\label{Dashen}
\left(\frac{\hat m + m_s}{\hat m}\right) \{1+\Delta_M\} = 
\frac{m^2_{K^0}+m^2_{K^+}-r_D (m^2_{\pi^+}-m^2_{\pi^0})}{m^2_{\pi^0}}\ , 
\end{equation}
where $2\hat m \equiv m_u+m_d$ and
$r_D \simeq 1.8 \pm 0.2$ accounts for the departure from the Dashen theorem value,
$r_D = 1$, of the photonic (self-energy) contribution to pseudoscalar mass
differences. From Eq.~(\ref{Dashen}) one gets $m_s/{\hat m} = 24.0 \pm 0.4$ to be
compared with $m_s/{\hat m} = 24.4 \pm 1.5$ from Ref.~\cite{LEU2}. An alternative
(but compatible) approach to Refs.~\cite{LEU,LEU2}, based on the tadpole scheme of
Coleman and Glashow \cite{CG}, has been recently discussed in Ref.~\cite{CS}. In that
case the Dashen theorem corrections are built in (properly explaining the $SU(2)$
electromagnetic mass splittings of ground state hadrons) with $r_D = 3/2$ in
Eq.~(\ref{Dashen}).

In summary, from the excellent and exhaustive sets of $J/\psi \to VP$ decay data
measured by the MARK III and DM2 detectors, one can extract a rather accurate value
for the $\eta$-$\eta\prime$ mixing angle using widely accepted models and well-tested
phenomenology. The value is found to be 
$\theta_P = -16.9^\circ \pm 1.7^\circ$, more than one standard deviation less
negative than all previous estimates. This difference is essentially due to the
introduction in our analysis of the corrections originated by the departure of
$\omega$-$\phi$ mixing from the ideal mixing considered by the other authors.

{\it Acknowledgements:}
Thanks are due to the grants CICYT-AEN95-815 and CIRIT-96SGR-21 for financial
support. M.~D.~S.~appreciates the hospitality at the Universitat Aut\`onoma de
Barcelona.

\tableone

\end{document}